\documentclass[%
 reprint,
 amsmath,amssymb,
 aps,
]{revtex4-2}

\usepackage{graphicx}
\usepackage{dcolumn}
\usepackage{bm}
\usepackage[utf8]{inputenc}
\usepackage{color}
\usepackage{graphics}

\usepackage{hyperref}
\usepackage{cleveref}
\usepackage{ifpdf} 
\usepackage{url}

\begin{document}

\preprint{APS/123-QED}

\title{Advanced light-shift compensation protocol in a\\ continuous-wave
microcell atomic clock}

\author{M. Abdel Hafiz$^1$, R. Vicarini$^1$, N. Passilly$^1$, C. E. Calosso$^2$, V. Maurice$^1$,\\ J. W. Pollock$^{3,4}$, A. V. Taichenachev$^{5,6}$, V. I. Yudin$^{5,6,7}$, J. Kitching$^3$ and R. Boudot$^{1}$}

\affiliation{$^1$FEMTO-ST, CNRS, UBFC, ENSMM, 26 rue de l'\'epitaphe 25030 Besan\c{c}on, France}
\affiliation{$^2$INRIM, Strada delle Cacce 91, Torino, Italy}
\affiliation{$^3$National Institute of Standards and Technology, Boulder, Colorado 80305, USA}
\affiliation{$^4$University of Colorado, Boulder, Colorado 80309-0440, USA}
\affiliation{$^5$Novosibirsk State University, ul. Pirogova 1, Novosibirsk 630090, Russia}
\affiliation{$^6$Institute of Laser Physics SB RAS, pr. Akademika Lavrent’eva 15B, Novosibirsk 630090, Russia}
\affiliation{$^7$Novosibirsk State Technical University, pr. Karla Marksa 20, Novosibirsk 630073, Russia}


\date{\today}

\begin{abstract}
Light-shifts are known to be an important limitation to the mid- and long-term fractional frequency stability of different types of atomic clocks. In this article, we demonstrate the experimental implementation of an advanced anti-light shift interrogation protocol onto a continuous-wave (CW) microcell atomic clock based on coherent population trapping (CPT). The method, inspired by the Auto-Balanced Ramsey (ABR) spectroscopy technique demonstrated in pulsed atomic clocks, consists in the extraction of atomic-based information from two successive light-shifted clock frequencies obtained at two different laser power values. Two error signals, computed from the linear combination of signals acquired along a symmetric sequence, are managed in a dual-loop configuration to generate a clock frequency free from light-shift. Using this method, the sensitivity of the clock frequency to both laser power and microwave power variations can be reduced by more than an order of magnitude compared to normal operation. In the present experiment, the consideration of the non-linear light-shift dependence allowed to enhance light-shift mitigation. The implemented technique allows a clear improvement of the clock Allan deviation for time scales higher than 1000~s. This method could be applied in various kinds of atomic clocks such as CPT-based atomic clocks, double-resonance Rb clocks, or cell-stabilized lasers.
\end{abstract}

\maketitle

\section{\label{sec:Intro}Introduction}
Atomic clocks have experienced over the last decades a remarkable evolution and progress. On the one hand, outstanding progress in laser science and technology, making possible manipulation, cooling, trapping, and probing of ultra-narrow atomic transitions, has led to the demonstration of optical atomic frequency standards with fractional frequency stability in the range of a few 10$^{-17}$ $\tau^{-1/2}$ and accuracy approaching the 10$^{-19}$ level \cite{Schioppo:Nature:2017, Ye:2019}. These instruments are exquisite tools to perform fundamental physics tests including geodesy \cite{Grotti:Nature:2018, McGrew:2018}, Lorentz symmetry testing \cite{Sanner:2019}, and dark matter searching \cite{Wcislo}. On the other hand, the combination of atomic spectroscopy, integrated photonics, and microelectromechanical systems (MEMS) has permitted the demonstration of low-power fully-miniaturized atomic clocks with stability levels below the 10$^{-11}$ level at 1 day averaging time \cite{Lutwak:PTTI:2007, Kitching:RevCSAD:2018, ULPAC, Vicarini:UFFC:2019}, finding applications in many civil and industrial systems.\\
For a wide variety of atomic clocks, including optical clocks based on quadrupole \cite{Huang:PRA:2012}, octupole \cite{Hosaka:PRA:2009} and two-photon transitions \cite{Gerginov:PRAp:2018, Newman:Optica:2018, Martin:PRAp:2018, Martin:PRA:2019}, microwave vapor cell clocks based on coherent population trapping (CPT) \cite{Hemmer:JOSAB:1989, Yano:PRA:2014, MAH:JAP:2017, Yun:PRAp:2017, Pollock:PRA:2018} or optical-microwave double-resonance technique \cite{Almat:UFFC:2020}, light-induced frequency shifts are known to be a relevant contribution to the clock mid- and long-term frequency stability performance.\\
A common approach to mitigate light-shifts is to probe the atomic resonance in pulsed operation using Ramsey spectroscopy \cite{Ramsey:PR:1950}. However, with Ramsey spectroscopy, interrogation pulses themselves are responsible for non-negligible light-shifts. Thus, advanced Ramsey interrogation protocols have been proposed and demonstrated that rely on the generation of two consecutive Ramsey sequences with different dark periods and the subsequent extraction of error signal(s) for the local oscillator (LO) frequency stabilization and light-shift compensation \cite{Sanner:PRL:2018, Yudin:PRAp:2018, Zanon:RPP:2018, Yudin:NJP:2018, MAH:PRAp:2018, MAH:APL:2018, Shuker:PRL:2019, Shuker:APL:2019}. Nevertheless, these methods are dedicated to pulsed-operation clocks and are not adapted for continuous-wave (CW) clocks that yet, still represent the privileged operation mode in numerous atomic clocks, including Rb cell clocks, chip-scale atomic clocks, or cell-stabilized lasers.\\
Therefore, the establishment of light-shift suppression techniques in CW clocks is of significant interest. In two-photon optical clocks, original techniques based on the use of two interrogating laser fields at different frequencies \cite{Gerginov:PRAp:2018}, the proper tuning of a magic light-field  polarization \cite{Jackson:PRA:2019} or the use of a dual-color magic wavelength trap \cite{Hilton:PRAp:2019} have been reported. In optically-pumped Rb clocks, a method of suppressing the light-shift by adjusting the laser frequency used for optical pumping was proposed in \cite{McGuyer:APL:2009}. In CPT atomic clocks, several approaches for light-shift mitigation have been proposed, including the fine tuning of the VCSEL microwave modulation index \cite{Zhu:Patent:2001, Shah:APL:2006}, the compensation for the laser frequency detuning \cite{Zhang:JOSAB:2016}, the adjustment of the cell temperature to a specific setpoint \cite{Miletic:APB:2012} or solutions combining actions on both the VCSEL dc-bias current drift and on the modulation index of the VCSEL light field \cite{Yanagimachi}. Nevertheless, the latter techniques rely on experimentally-determined quantities that are expected to change with time and from one device to another. \\
In a recent study, novel universal concepts have been proposed for compensating the light-shift and its fluctuations in CW-atomic clocks  \cite{Yudin:ArXiv:2019}. In one of the proposed methods, named CW-Auto Compensated Shift (CW-ACS), the LO frequency is successively adjusted at two light-shifted frequencies obtained at two distinct laser powers through a control parameter. Assuming linearity of the light-shift, the latter is then used as an artificial anti-light shift to compensate for the actual undesired light-shift. However, to our knowledge, these techniques have never been demonstrated experimentally.\\
In this article, we demonstrate the experimental implementation of the CW-ACS interrogation protocol applied to a CPT-based microcell atomic clock. Inspired by \cite{MAH:APL:2018}, the CW-ACS sequence is symmetrized (CW-SACS) for cancellation at the first order of an atomic memory effect and to compensate for errors in the atomic frequency estimation induced by imperfections of the laser power modulation pattern. Original error signals are processed such that all pulses are used both for the LO frequency and light-shift correction servo loops, preventing the presence of a dead time and the loss of information. Moreover, in the reported CPT clock, an extended version of the CW-SACS technique was applied to overcome the non-linear response of the light-shift curve and then improve the light-shift reduction. Using this approach, the sensitivity of the clock frequency to laser and microwave power variations is further reduced by more than a factor 10. The implemented technique allows a clear improvement of the clock Allan deviation for time scales higher than 1000 s.

\section{\label{sec:Basics}Basics of the CW-ACS method}

In the present manuscript, we consider for illustration a Cs atomic system, with unperturbed clock transition frequency $\nu_{Cs}$, that can be affected by different kinds of frequency shifts. The CW-ACS method, proposed in \cite{Yudin:ArXiv:2019}, is an attractive approach in atomic systems where light-induced frequency shifts are dominant. The main objective of the CW-ACS method is to generate a clock frequency equal to the light-shift-free frequency $\nu_{at}$.\\
Let's consider first an ideal light-shift free passive atomic clock. From an experimental point of view, the LO frequency is first swept in order to detect the atomic resonance signal, characterized by its linewidth $\Delta \nu$. Hence, a dispersive error signal $S$, crossing zero at the resonance extremum, is obtained by probing, with $\pm \Delta \nu /2$ frequency-jumps of the LO, the signal on both sides of the atomic resonance. An iterative algorithm servo loop is then implemented to null the error signal $S$ and then stabilize the LO frequency $\nu_{LO}$ onto the light-shift free atomic resonance $\nu_{at}$ such that $S (\nu_{LO}) \simeq - D_e (\nu_{LO} - \nu_{at}) = 0$, with $D_e$ the absolute value of the slope of the error signal $S$, at low detunings.\\
Under interaction with a light field, atoms experience a light-shift $\Delta_P$. The zero-crossing of the error signal is therefore shifted by $\Delta_P$, leading the LO frequency to be stabilized onto a light-shifted frequency, since the servo loop operates to satisfy $S (\nu_{LO}) \simeq - D_e (\nu_{LO} - (\nu_{at} + \Delta_P)) = 0$. Moreover, since this signal depends on the laser power, laser power variations will induce frequency instabilities. Let's consider now two light-shifted error signals $S_1$ and $S_2$ obtained at two laser powers $P_1$ and $P_2$, such that $P_2 > P_1$. Linearity of the light-shift is assumed such that the actual light-shift at the power $P_1$ ($P_2$) is $\Delta_{P_1} = c P_1$ ($\Delta_{P_2} = c P_2$), with $c$ the coefficient of proportionality. The essence of the CW-ACS method is then for any laser power $P$ to shift the frequency sent to the atoms $\nu_{LO}$ from the unperturbed resonance frequency by an amount $\xi P$ so that $\nu_{LO}= \nu_0 + \xi P$, where $\nu_0$ and $\xi$ are two variable control parameters that, respectively, estimate the light-shift free frequency $\nu_{at}$ and the actual light-shift $cP$, such that $\xi P - cP = 0$. In an operation sequence where the laser power is alternated between two values $P_1$ and $P_2$, the output result is equivalent to a system of two equations with two unknown variables $\nu_{0}$ and $\xi$ such that the two error signals are:
\begin{align}
S_1 (\nu_{0} + \xi P_1) = - D_{e_1}~(\nu_{0}+\xi P_1-\nu_{at}-c P_1) = 0\\
S_2 (\nu_{0} + \xi P_2) = - D_{e_2}~(\nu_{0}+\xi P_2-\nu_{at}-c P_2) = 0
\end{align}
with $D_{e_i}$ the slope of the error signal at the power $P_i$, with $i=1,2$.
Solutions to this system are given by the relations:
\begin{align}
\nu_{0} &= \nu_{at} \\
\xi &= c
\end{align}
establishing that the CW-ACS method permits stabilization of the frequency $\nu_0$ onto the light-shift free atomic frequency $\nu_{at}$ and to determine the amount of actual light shifts $\Delta_{P_1} = c P_1$ and $\Delta_{P_2} = c P_2$. 

\section{Symmetric CW-ACS sequence}
\begin{figure*}[t]
\centering
\includegraphics[width=0.7\linewidth]{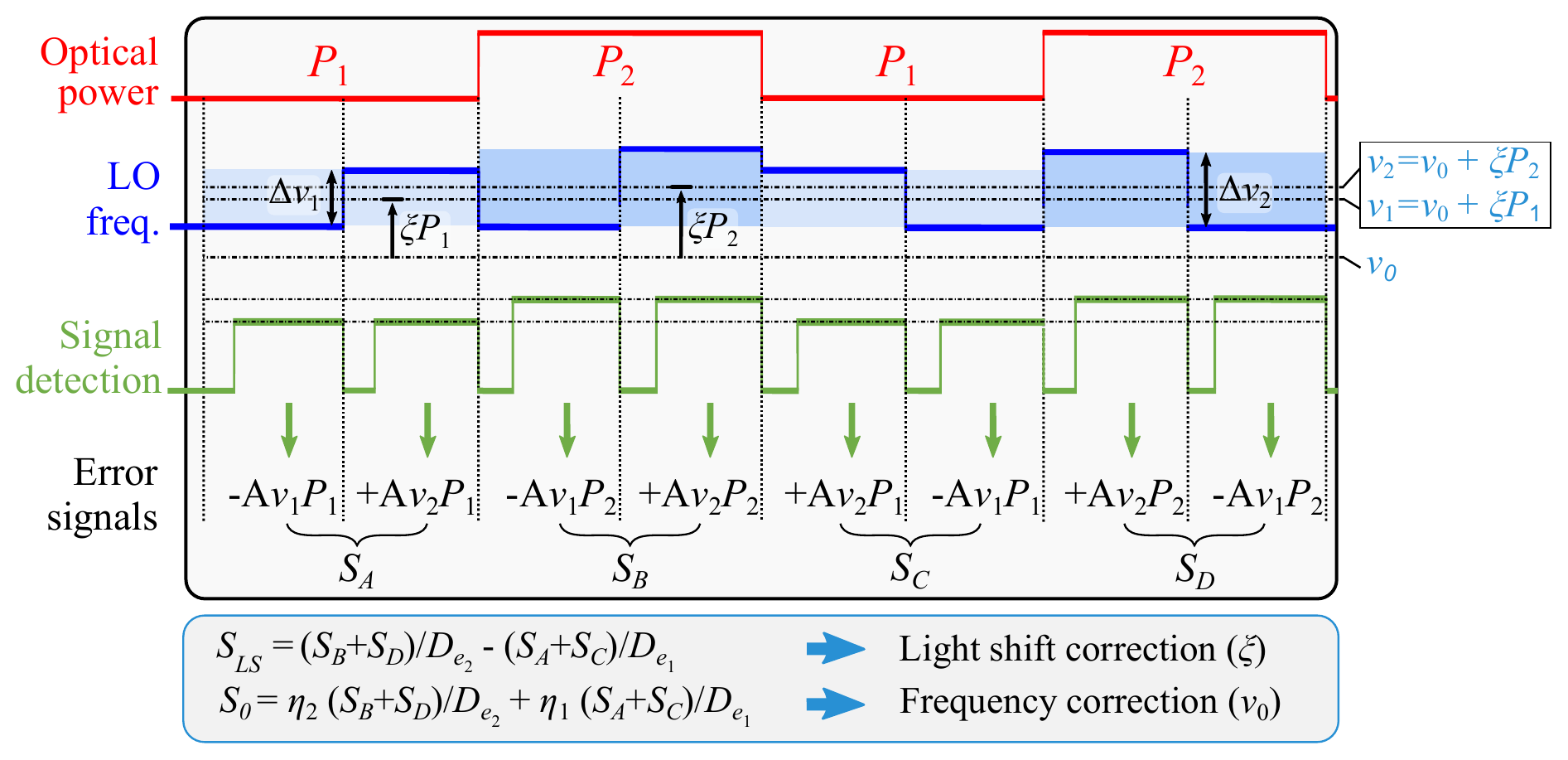}
\caption{Symmetric CW-ACS sequence applied to a microcell CPT-based atomic clock. $A\nu_iP_i$, with $i=1,2$, denote the respective atomic signal values acquired at half-height of the resonance along the sequence. Sign ($+$ or $-$) attached to the signal values $A\nu_iP_i$ are those considered to calculate the error signals $S_A$ to $S_D$. In our experiment, the LO modulation frequency is 1.45 kHz. The photodiode signal is acquired (signal detection) after a delay of 70 $\mu$s at each probe frequency change. This delay corresponds roughly to the response time of the laser frequency servo.} 
\label{fig:Sequence}
\end{figure*}

Figure \ref{fig:Sequence} illustrates the CW-SACS interrogation sequence we applied to a CPT-based microcell atomic clock, where $A\nu_iP_i$, with $i=1,2$, denotes the averaged atomic signal over a step with LO frequency $\nu_i$ and laser power $P_i$.
This sequence is divided into two CW-ACS sub-sequences. In each sub-sequence, the laser power is successively tuned at values $P_1$ and $P_2$, such that $P_2$ $>$ $P_1$. In the first sub-sequence, the CPT signal detection is performed by tuning first the LO frequency $\nu_{LO}$ at $\nu_0 + \xi P_1 \mp \Delta \nu_1/2$ (the left then the right side of the resonance) with light power $P_1$, and then at $\nu_0 + \xi P_2 \mp \Delta \nu_2/2$ with power $P_2$. The applied modulation depths equal the CPT resonance linewidth $\Delta \nu_1$ ($\Delta \nu_2$) measured at the power $P_1$ ($P_2$). In the second sub-sequence, the detection is performed by tuning first the LO frequency at $\nu_0 + \xi P_1 \pm \Delta \nu_1/2$ (at the right then the left side of the resonance) with light power $P_1$, then at $\nu_0 + \xi P_2 \pm \Delta \nu_2/2$ with power $P_2$.\\
In the first (second) sub-sequence, two successive error signals $S_A$ and $S_B$ ($S_C$ and $S_D$) are extracted for power values $P_1$ and $P_2$, respectively. 
In comparison with the CW-ACS proposal reported in \cite{Yudin:ArXiv:2019}, an important difference is the use of an extended symmetric sequence. The CW-ACS sequence is symmetrized first to cancel at the first order a memory effect of the atoms. This memory effect, already highlighted in the case of a pulsed Ramsey-CPT clock \cite{MAH:APL:2018}, is explained here from a qualitative point of view. The key point is to understand that the atomic signal detected at a given time $t$, instead of depending only on the actual probing frequency at this time $t$, carries also, for a period equivalent to the population and CPT coherence relaxation times \cite{Boudot:JOSAB:2018}, some information related to preceding levels of the interrogation microwave and laser fields. In the present experiment, the population of atoms in the CPT state is higher at $P_2$ than at $P_1$ and it takes for this population up to a few milliseconds to stabilize after a change in the laser power set point. This results in a deviation of the atomic signal from the one obtained when no power modulation is applied. The deviation is enforced (reduced) if, during the previous step, the microwave frequency is on the same (opposite) side of the atomic resonance with respect to the current one. Considering the first half of the total symmetric sequence shown in Fig. 1, one expects then $A\nu_1P_1$ to be slightly positively offset and $A\nu_1P_2$ negatively offset. The value of the detected atomic signal $S_A$ is then overestimated and the value of $S_B$ underestimated. This error is then compensated at the first order with the second half of the sequence shown in Fig. \ref{fig:Sequence} during which the LO frequency modulation pattern is the mirror symmetric to the one used in the first half of the sequence. The second reason to use a symmetric sequence is that the laser power modulation pattern, generated in this manuscript using an acousto-optic modulator (AOM), exhibits experimentally slight transient-type imperfections. The atomic signal detected at the output of the cell is then polluted by spurious voltage offsets that induce frequency offsets in the estimation of the atomic resonant frequency. \\
Successive errors signals ($S_A$ to $S_D$) acquired along the sequence are combined to produce two output error signals. The first output error signal $S_{LS}$, such that:
\begin{equation}
S_{LS} = \frac{S_B + S_D}{D_{e_2}} - \frac{S_A + S_C}{D_{e_1}} \simeq - 2 (\xi-c) (P_2-P_1)
\label{eq:sls}
\end{equation}
is equivalent to a weighted difference of (1) and (2) and is representative of the residual light-shift. It is nulled by an integrative controller that corrects the control parameter $\xi$, in order to compensate for the light-induced frequency shift. Once the light-shift is compensated, a second output error signal $S_{0}$, defined by;
\begin{equation}
S_{0} = \eta_2 \frac{S_B + S_D}{D_{e_2}} + \eta_1 \frac{S_A + S_C}{D_{e_1}} \simeq -2 (\nu_{0} - \nu_{at})
\label{eq:s0}
\end{equation}
is calculated and is used by a second integrative controller for correcting the base frequency $\nu_{0}$, so that it converges to the light-shift free frequency $\nu_{at}$. $\eta_1$ and $\eta_2$
are two weighting coefficients ($\eta_1 + \eta_2 =$ 1) that can be used to improve the short-term stability if they are inversely
proportional to the short-term stability of the clock at
powers $P_1$ and $P_2$, respectively \cite{Tavella:1994}.\\
The use of the combination of Eqs (\ref{eq:sls}) and (\ref{eq:s0}) is advantageous for two main reasons. The first reason is that all error signals ($S_A$, $S_B$, $S_C$, $S_D$) are used in both servo loops, preventing the presence of a dead time and the consequent loss of information through aliasing \cite{Calosso:2020}. In this way, the noise in the estimations is reduced, with beneficial effects on the short-term stability of the clock. The latter, is further optimized by the weighting coefficients $\eta_1$ and $\eta_2$, which allow to combine optimally the short-term stability of the clock at different laser powers, in case they are different. The second reason is that $S_{0}$ is a direct representative of the difference $\nu_0 - \nu_{at}$ in closed-loop operation, while $S_{LS}$ carries directly the residual light-shift information, in any condition and independently of $\nu_0$. This configuration shows high decoupling of the two loops and prevents them to interfere with each other. It is interesting to note that the normalization through $D_{e_1}$ and $D_{e_2}$ and weighting do not change the solution of Eqs (1) and (2) and the knowledge of their exact value is not critical like in other methods such as in the CW-CES method presented in \cite{Yudin:ArXiv:2019}. Nevertheless, their use is important from an experimental point of view, since (5) and (6) allow to consider and optimize noise and robustness, whereas these aspects are not usually taken into account by this kind of methods.

\section{Experimental set-up}
\begin{figure}[t]
\centering
\includegraphics[width=\linewidth]{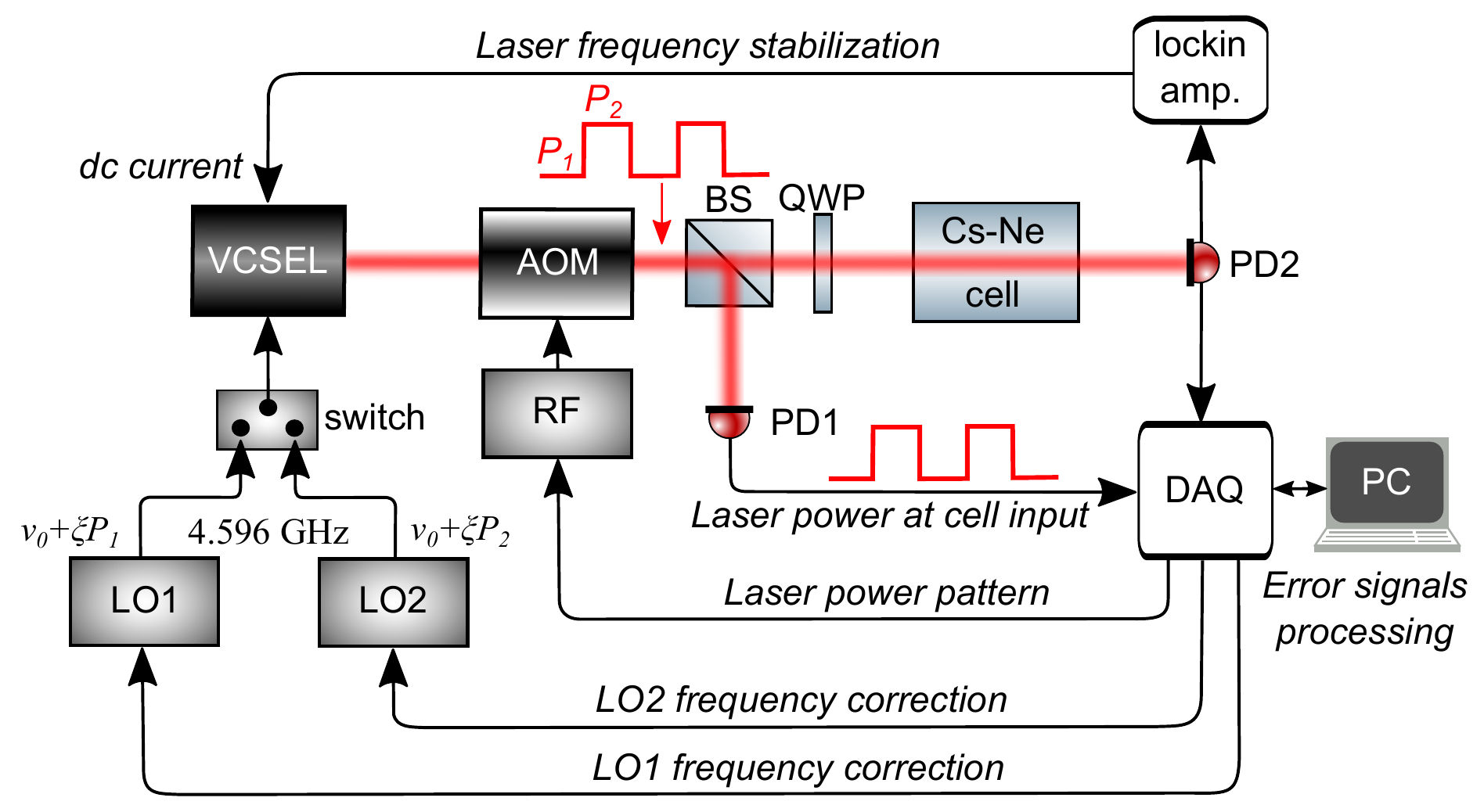}
\caption{CPT clock experimental setup for test of the CW-SACS method.}
\label{fig:setup-cw-acs}
\end{figure}

The symmetric CW-ACS sequence was tested in the CPT-based microcell atomic clock described in Fig. \ref{fig:setup-cw-acs}. 
The laser source is a vertical-cavity surface emitting laser (VCSEL) tuned on the Cs D$_1$ line at 894.6 nm \cite{Kroemer:AO:2016}. The injection current of the VCSEL is directly microwave-modulated at 4.596 GHz in order to produce two first-order optical sidebands separated by 9.192 GHz for the CPT interaction. An AOM, driven by a radiofrequency (RF) synthesizer, is used to produce laser power modulation and generate consecutive laser powers $P_1$ and $P_2$. At the output of the AOM, a fraction of the laser power is reflected by a beam splitter and detected by a photodiode (PD1) before the cell input. The transmitted laser beam is circularly polarized using a quarter-wave plate (QWP) to produce the CPT interaction in the vapor cell. The heart of the atomic clock is a Cs microfabricated cell \cite{Hasegawa:SA:2011}, filled with a Ne buffer gas pressure of about 93 Torr \cite{Kozlova:PRA:2011}. The CPT interaction takes place in a cavity of the cell with a diameter of 2 mm and a length of 1.4 mm. The cell is temperature-stabilized at about 82$^{\circ}$C for reduced sensitivity of the clock frequency to cell temperature variations \cite{Miletic:EL:2010, Kozlova:TIM:2011}. A static quantization magnetic field is produced in order to raise the Zeeman degeneracy and isolate the 0-0-clock transition. A mu-metal shield is used to protect the cell from external magnetic field perturbations. The laser power transmitted through the cell is detected by a photodiode (PD2) from which the atomic signal information is extracted. In the first path, this signal is used with the help of a lock-in amplifier for laser frequency stabilization onto the bottom of the absorption profile ($F'$=4 excited state). In the second path, the signal is acquired by a I/O multi-function card. The latter manages the CW-SACS sequence generation as well as the extraction of the error signals for frequency stabilization of the clock on the light-shift free frequency. \\
In the paper proposing the CW-ACS method \cite{Yudin:ArXiv:2019}, the use of an additional electronic block (frequency shifter) able to shift the LO frequency for any laser power by the correct light-shift value was suggested. In our setup, the real-time management with a single LO block of the CW-SACS sequence was challenging due to the limited performance of our multi-function card. Thus, we chose for demonstration to alternately drive the VCSEL with two commercial frequency synthesizers, referenced by the same source, using a high-isolation (55 dB at 4.6 GHz) broadband switch as a microwave multiplexer. The first (second) synthesizer generates the shifted frequency $\nu_1 = \nu_0 + \xi P_1$ ($\nu_2 = \nu_0 + \xi P_2$) and associated frequency jumps $\pm \Delta \nu_1/2$ ($\pm \Delta \nu_2/2$) to probe the CPT resonance at the power $P_1$ ($P_2$). Frequency values of both LO synthesizers are then kept resonant with the light-shifted atomic frequencies, in agreement with their respective individual power values. The light-shift free atomic frequency $\nu_0$ is then simply computed from frequencies $\nu_1$ and $\nu_2$. Note that for simplicity of experiments, we chose to apply frequency-jumps $\Delta \nu_1$= $\Delta \nu_2$= 2.5 kHz, whatever the laser power. The servo gains were adapted to have roughly the same servo bandwidth for all experiments.
\begin{figure}[t]
\centering
\includegraphics[width=\linewidth]{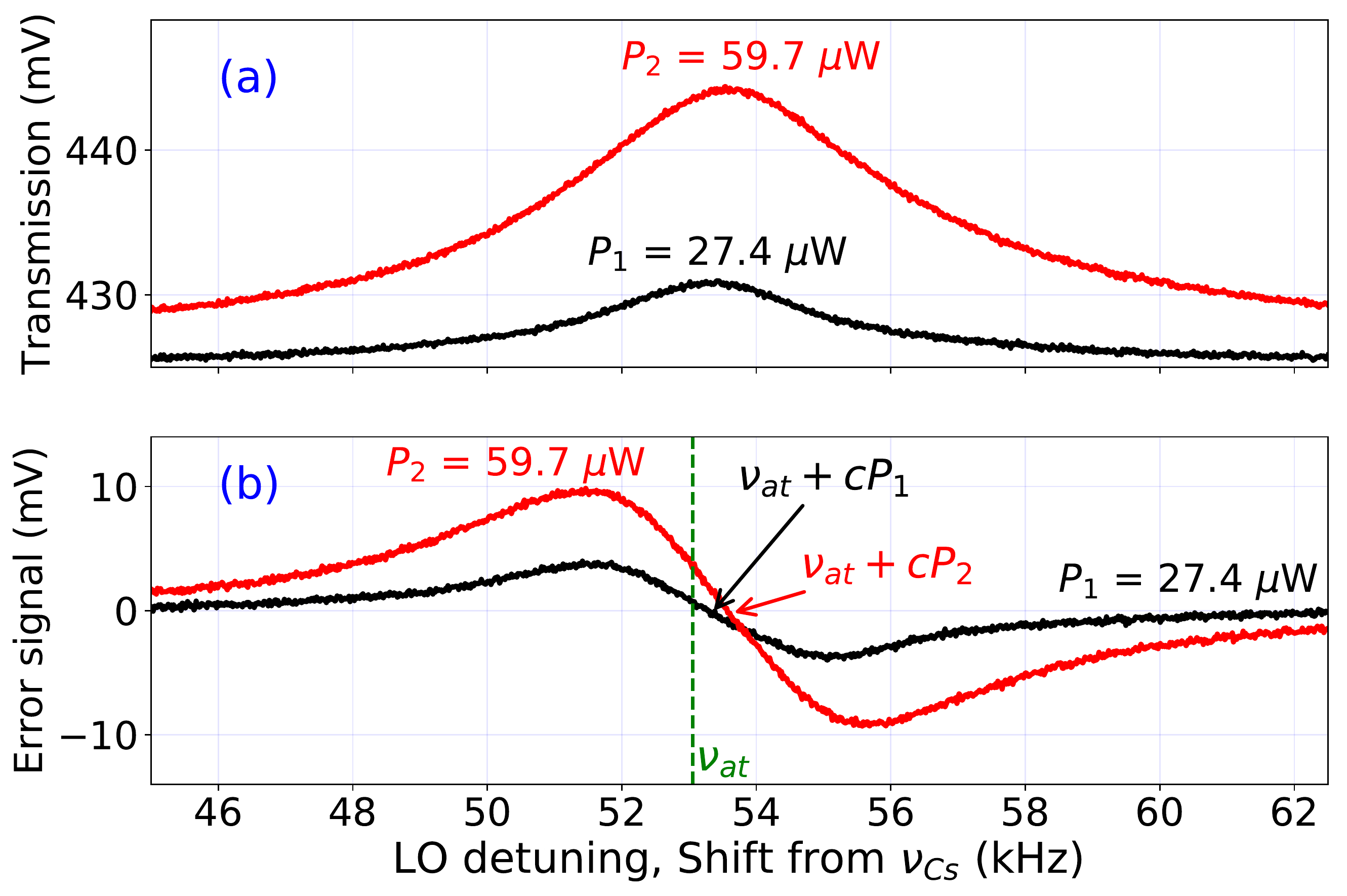}
\caption{(Color online) CPT resonances (a) and associated error signals (b) obtained at two different laser powers $P_1 =$ 27.4 $\mu$W and $P_2$ = 59.7 $\mu$W. An offset of 0.51 V was subtracted from the CPT resonance measured at $P_2$ for clarity. The green vertical dashed line shows the approximate position of the light-shift free frequency $\nu_{at}$. Light-shifted frequencies $\nu_{at} + c P_1$ and $\nu_{at} + c P_2$, identified by the zero-crossing of the error signals, are indicated by arrows.}
\label{fig:spectra}
\end{figure}

\section{Experimental results}
\begin{figure}[t]
\centering
\includegraphics[width=0.9\linewidth]{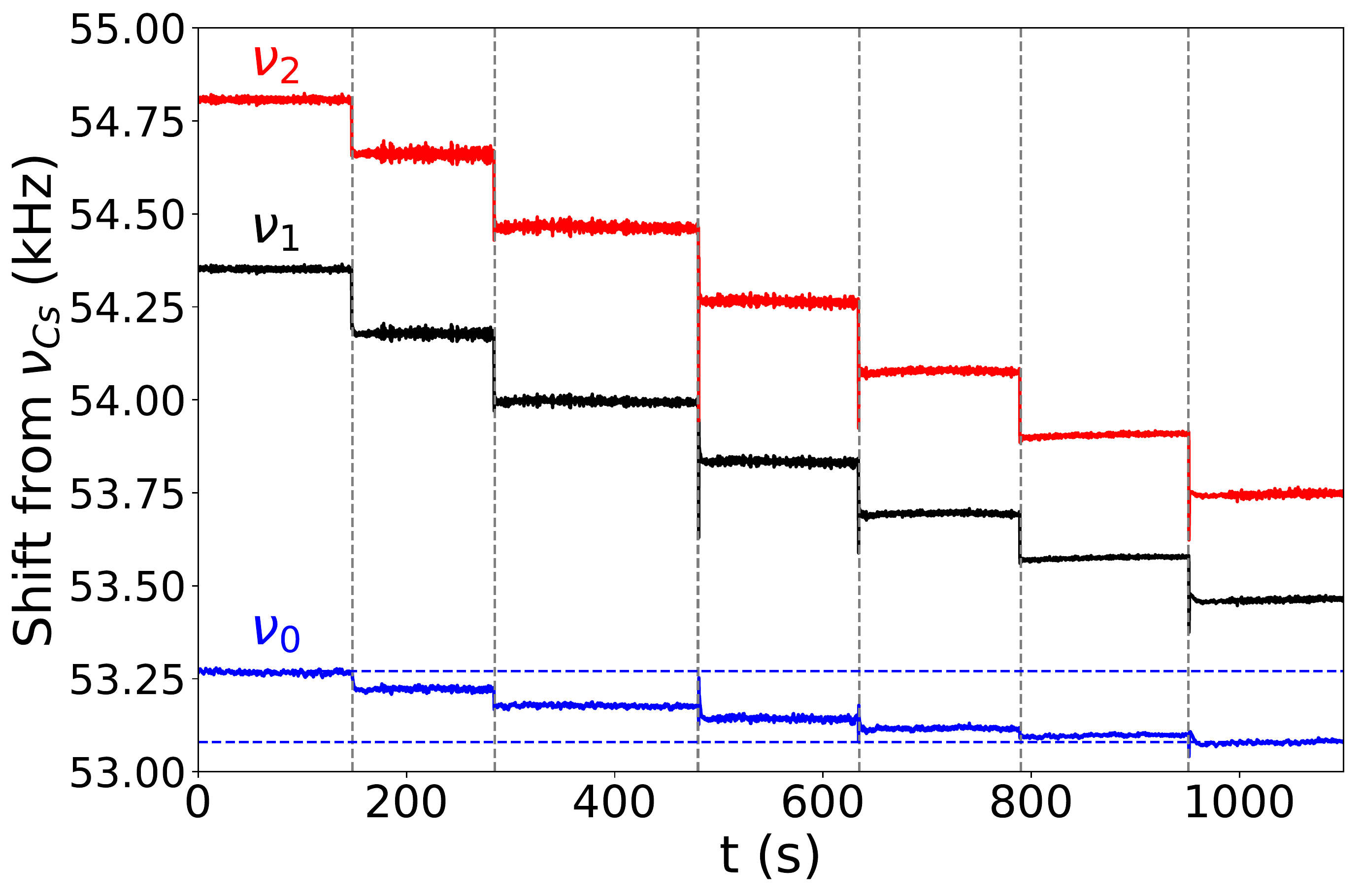}
\caption{Measurement of frequencies $\nu_1$, $\nu_2$, and $\nu_{0}$ (frequency shift relative to $\nu_{Cs}$) in a clock run using the symmetric CW-ACS sequence. Each power jump is shown by a vertical dashed line. For the total span, the power $P_1$ ($P_2$) is changed from 80.5 to 25 $\mu$W (114 to 43.5 $\mu$W). Each power jump corresponds to a 1 dB variation on the RF synthesizer driving the AOM used to control the laser power. The microwave power $P_{\mu}$ is 2.17 dBm. The $\pm$ 2.5 kHz frequency-jumps are applied for each power $P_1$ and $P_2$ to scan the CPT resonances. The power modulation depth is 35 \%.}
\label{fig:SeqTemporelle}
\end{figure}

Figure \ref{fig:spectra} shows two CPT resonances and their associated error signals $S_1$ and $S_2$ obtained at respective power values $P_1=$  27.4 $\mu$W and $P_2=$ 59.7 $\mu$W, for a microwave power $P_{\mu}$ of 4.17 dBm. At the power value $P_1$ ($P_2$), the CPT resonance linewidth is 4.4 (6.5) kHz and the CPT resonance contrast is 1.2 \% (3.9 \%). The slope of the CPT error signal is in absolute value 3.0 and 6.8 $\mu$V/Hz, for power values $P_1$ and $P_2$ respectively. For both powers $P_i$, with $i=1,2$, the totally shifted clock frequency, identified by the zero-crossing of the error signal, is $\nu_{Cs}+ \Delta_Z + \Delta_{bg} + \Delta_{P_i} = \nu_{at} + c P_i$, where $\Delta_Z$ and $\Delta_{bg}$, both independent of  the laser power, are respectively the Zeeman shift ($\sim$ 4.27 Hz) and the buffer gas-induced collisional frequency shift \cite{Kozlova:TIM:2011}, and $\Delta_{P_i} = c P_i$ is the light-shit. From Fig. 3, we find a light-shift ratio  $\Delta_{P_2}/\Delta_{P_1}$ $\sim$ 570 Hz/274 Hz $\sim$ 2.08 $\pm$ 0.06 for a power ratio $P_2/P_1$ $\sim$ 2.18 $\pm$ 0.01. The slight discrepancy between these two ratios can be explained by the actual non-linearity of the light-shift dependence, as highlighted later in the manuscript. \\
Figure \ref{fig:SeqTemporelle} shows a temporal plot in clock operation using the symmetric CW-ACS sequence. This test is performed with a microwave power of 2.17 dBm. During the clock run, light-shifts induced by sudden laser power jumps (vertical dashed grey lines) are applied while frequencies $\nu_1$, $\nu_2$, and $\nu_0$ are recorded. A significant variation of both frequencies $\nu_1$ and $\nu_2$ is observed at each laser power change whereas the value of the frequency $\nu_0$ is less affected by laser power variations. In this test, the total $P_2$ power change of 70.5 $\mu$W induces a 1060 Hz change of the frequency $\nu_2$, yielding a sensitivity of about 15 Hz/$\mu$W, i.e. 1.6 $\times$ 10$^{-9}$/$\mu$W in fractional value. This sensitivity is similar for the frequency $\nu_1$, whereas it is reduced to the level of about 3 Hz/$\mu$W for the frequency $\nu_0$. This result demonstrates a reduction of the light-shift coefficient (LSC) by about a factor 5 in the CW-SACS mode.\\
\begin{figure}[t]
\centering
\includegraphics[width=\linewidth]{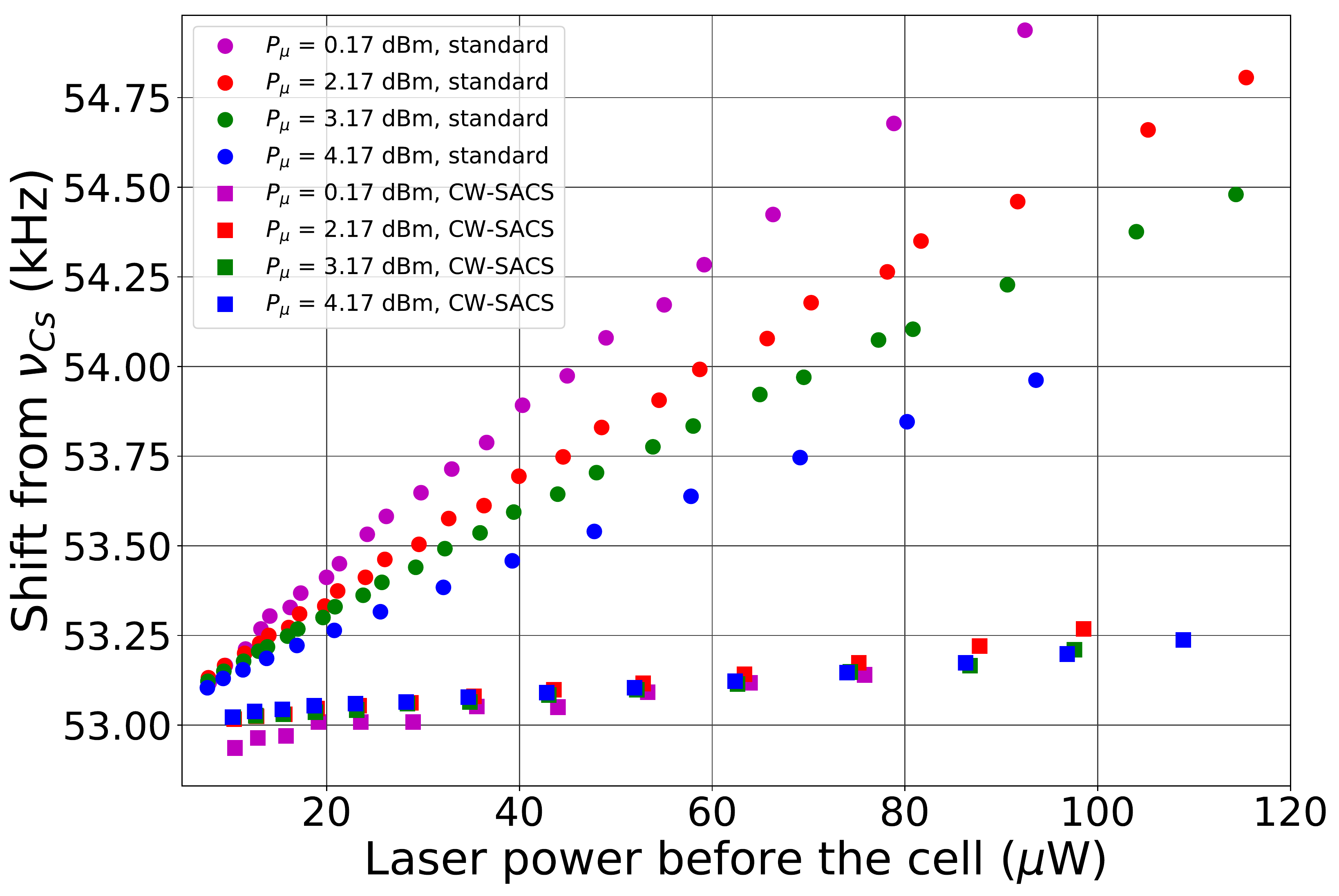}
\caption{Clock frequency shift (referenced to $\nu_{Cs}$) versus the laser power at the cell input, for different microwave powers (0.17, 2.17, 3.17, and 4.17 dBm) driving the VCSEL, in standard (circles) or CW-SACS operation (squares). For the CW-SACS case, the average power $(P_1 + P_2)/2$ is considered. Error bars are smaller than the data points symbols.}
\label{fig:lightshift_Normal_ACS_35pc_Fmdev2500Hz}
\end{figure}
Extracted from sequences similar to the one shown in Fig. \ref{fig:SeqTemporelle}, Fig. \ref{fig:lightshift_Normal_ACS_35pc_Fmdev2500Hz} shows the total clock frequency shift (referenced to the unperturbed Cs frequency $\nu_{Cs}$= 9.192631770 GHz) versus the laser power at the cell input, for different microwave powers $P_{\mu}$, in the standard CW interrogation scheme (without any light-shift mitigation technique), in comparison with the symmetric CW-ACS scheme. In the standard mode, the LSC decreases with increased microwave power. It is about 22 Hz/$\mu$W and 10 Hz/$\mu$W for microwave powers of 0.17 dBm and 4.17 dBm, respectively. For modulation powers higher than 4.17 dBm, we found that the CPT contrast and then the clock short-term frequency stability started to degrade due to reduction of the power contained in CPT-resonant first-order sidebands. The short-term stability of the CPT clock being optimal for laser powers between 20 $\mu$W and 60 $\mu$W, we focused the following analysis in this range. In the CW-SACS mode, the residual LSC in the 20-60 $\mu$W power range is about 2 Hz/$\mu$W, independent of the microwave power value. This sensitivity is 11 and 5 times smaller than the one obtained in the standard operation mode, at $P_{\mu}$ = 0.17 dBm and 4.17 dBm, respectively. This result shows firstly that the CW-SACS method is attractive to obtain light-shift mitigation, independently of the laser power working point. In addition, we find from Fig. \ref{fig:lightshift_Normal_ACS_35pc_Fmdev2500Hz} that the CW-SACS method reduces at fixed laser power the impact of microwave power variations on the clock frequency by about one order of magnitude.\\
Looking at Fig. \ref{fig:lightshift_Normal_ACS_35pc_Fmdev2500Hz}, which shows light shift measurement in a wider power range, a deviation of the power light-shift curve from a linear behaviour is observed. Such a non-linear light-shift behaviour, confirmed by plotting residuals obtained from the linear fit of experimental data points (not shown here), has already been reported in CPT spectroscopy \cite{Zhu:Patent:2001, Pollock:PRA:2018}. The rigorous theoretical explanation of this behaviour is not treated here. However, we stress that this deviation from the ideal linear light-shift response causes two main limitations to the CW-ACS method discussed here. The first one is the presence of a finite residual light-shift sensitivity contributing to degrade the clock stability. The second one is the presence of a finite clock frequency shift from the unperturbed transition frequency which depends on the absolute values of $P_1$ and $P_2$ and on their mutual difference.
\begin{figure}[t]
\centering
\includegraphics[width=\linewidth]{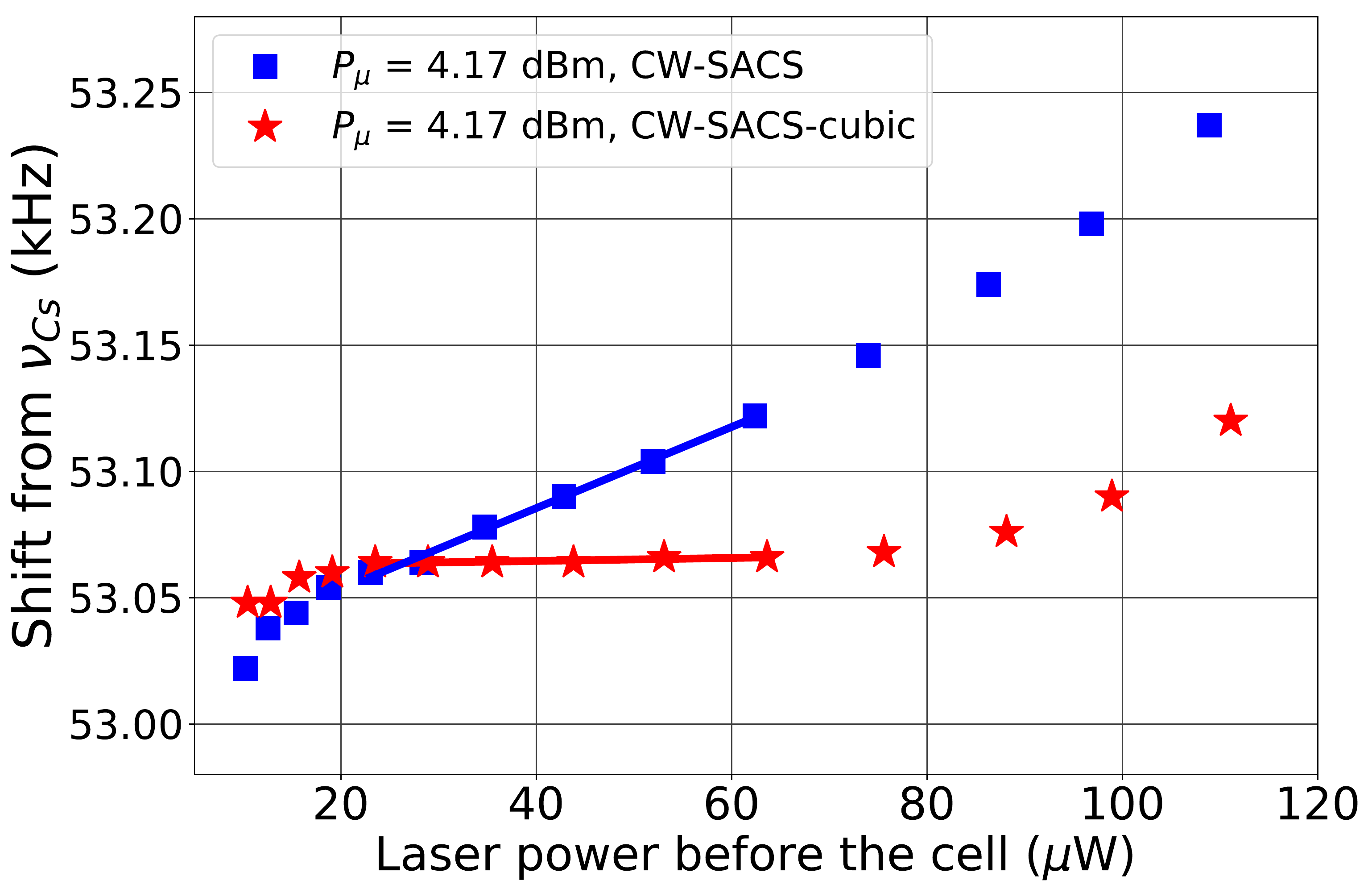}
\caption{Clock frequency shift (referenced to $\nu_{Cs}$) versus the laser power at the cell input using CW-ACS with two different approximations (squares: linear, stars: cubic) for the light-shift dependence. The microwave power is $P_{\mu}$=4.17 dBm. Error bars are smaller than the data points symbols. Solid lines indicate a linear fit to the data points, in the 20-63~$\mu$W range.}
\label{fig:linVsCubic}
\end{figure}
Thus, it is necessary to account for the light-shift non-linearity in order to improve the light-shift mitigation efficiency.\\
For this purpose, we extended the CW-ACS sequence algorithm by performing in a limited power range (20-63 $\mu$W) a third-order polynomial fit of experimental data. Definitions of $\Delta_{P_1} = \xi P_1$ and $\Delta_{P_2} = \xi P_2$ were therefore changed to $\xi P_1 + \beta P_1^2 + \gamma P_1^3$ and $\xi P_2 + \beta P_2^2 + \gamma P_2^3$, at powers $P_1$ and $P_2$, respectively. The second- ($\beta$) and third-order ($\gamma$) coefficients were extracted from the cubic fitting of experimental data and kept fixed during clock operation while $\xi$ is continuously calculated by the light-shift servo loop.\\
Figure \ref{fig:linVsCubic} displays the total clock frequency shift (from the Cs atom frequency $\nu_{Cs}$) versus the cell input laser power in the CW-SACS mode, for a microwave power of 4.17 dBm, when a linear or a cubic approximation is performed. In the 20-63 $\mu$W power range, the use of the cubic approximation reduces the LSC at the level of 0.059 Hz/$\mu$W, i.e. 6.4 $\times$ 10$^{-12}$$/$$\mu$W in fractional value. This value is 28 times smaller than with the CW-SACS method using the linear approximation of the light-shift dependence, and a factor 170 smaller than the one obtained in standard operation in the same conditions. We insist that the cubic-law approach can be employed here since we focus on demonstrating insensitivity to light-shifts but are not concerned with clock frequency accuracy.\\
Finally, we have performed measurements of the clock Allan deviation in the CW-SACS mode with the cubic function approximation of the light-shift dependence. Values of $\nu_1$, $\nu_2$, and $\nu_0$ were recorded throughout the whole sequence. The microwave power is 4.17 dBm. The clock is operated with power values $P_1=$ 28.4 $\mu$W and $P_2=$ 62.0~$\mu$W.
\begin{figure}[t]
\centering
\includegraphics[width=\linewidth]{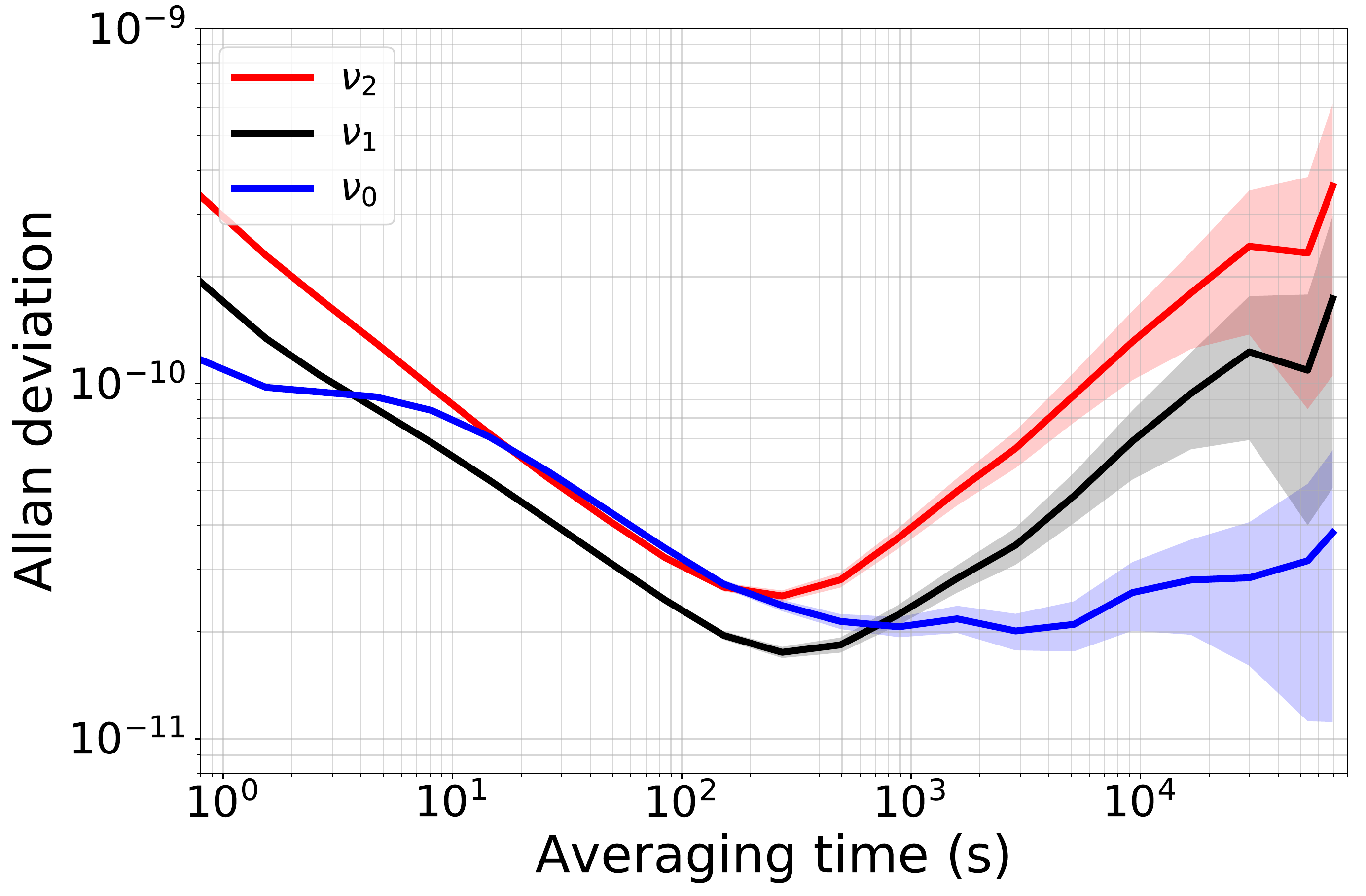}
\caption{Allan deviation of the frequencies $\nu_1$, $\nu_2$, and $\nu_0$ with CW-ACS using the cubic law approximation. Data are extracted from a measurement of 58 h and 33 minutes. Error bars are illustrated by lighter-colored zones.}
\label{fig:Adev}
\end{figure}
We used $\eta_1$ = $\eta_2$ = 0.5 and no specific short-term stability optimization was performed by tuning the values of these coefficients. The short-term stability of the frequency $\nu_1$, measured to be about 1.8~$\times$ 10$^{-10}$ $\tau^{-1/2}$ until 200 s, is better than that of $\nu_2$ (3 $\times$ 10$^{-10}$ $\tau^{-1/2}$). This is due to the fact that $P_1$ is closer than $P_2$ to the laser power (about 35 $\mu$W) that optimizes the clock short-term stability. In comparison with the standard clock configuration (single LO, no light-shift mitigation applied, laser power = 35 $\mu$W), we observed a degradation in the short-term stability by a factor of 4 for the frequency $\nu_1$. This degradation is partly explained by the power values choice mentioned above. In addition, sudden changes in optical power possibly degrading the laser frequency stabilization and the unusual simultaneous operation of two synthesizers might have contributed. Assuming the same fractional stability of the laser power at $P_1$ and $P_2$ and the same clock frequency sensitivity of about 10 Hz/$\mu$W for both power values, the medium-term stability ($\tau >$ 100 s) of $\nu_1$ is expected to be about 0.46 times better than the stability of $\nu_2$, since $P_1$ = 0.46 $P_2$. The plot shown in Fig. \ref{fig:Adev} is in agreement with this prediction and the mid-term stability of $\nu_1$ and $\nu_2$ is limited by the stability of the laser power.\\
The short-term stability of the frequency $\nu_{0}$ is similar to that of $\nu_2$ from 10 to 200 s. The bump at 10 s is attributed to the light-shift servo loop time constant. For timescales higher than 1000 s, the stability of the light-shift free frequency $\nu_0$ is clearly better than that of the light-shifted frequencies $\nu_1$ and $\nu_2$, with a stability level remaining in the 2-3 $\times$ 10$^{-11}$ range until 6 $\times$ 10$^4$~s. This result demonstrates further that the CW-SACS technique helps to mitigate the contribution of light-shift effects to the frequency stability of atomic clocks.

\section{Conclusions}
In conclusion, we have experimentally demonstrated a technique to mitigate light-shifts in a CW-CPT-based microcell atomic clock using a symmetric CW-ACS interrogation protocol \cite{Yudin:ArXiv:2019}. Two output error signals, constructed from the linear combination of consecutive signals acquired during the power-modulation based sequence, are used in a dual-loop system to generate a clock frequency free from light-shift. The clock frequency sensitivity to laser and microwave power variations can be reduced by one order of magnitude using the CW-SACS sequence. Improved mitigation was achieved in our CPT clock by approximating the light-shift dependence with a third-order polynomial function in the power range of interest. Tested on a similar clock setup and similar environmental conditions, the CW-SACS technique using cubic fitting of the light-shift dependence improved the clock Allan deviation for timescales higher than 1000 s. A more accurate understanding of light-shift mechanisms with an analytical expression that fits better to experimental data will certainly help to further improve the efficiency and robustness of this method.

\begin{acknowledgments}
The authors thank Azure Hansen and Joe Christensen (NIST) for careful reading of the manuscript before submission. The authors acknowledge Moshe Shuker for fruitful discussions on light-shift suppression techniques. This work was supported in part by D\'el\'egation G\'en\'erale de l'Armement, in part by Région de Franche-Comté, and in part by Agence Nationale de la Recherche (ANR) in the frame of the LabeX FIRST-TF (Grant ANR 10-LABX-0048), EquipX Oscillator-IMP (Grant ANR 11-EQPX-0033) and ASTRID PULSACION (Grant ANR-19-ASTR-0013-01) projects. V.~I.~Yudin was supported by the Russian Foundation for Basic Research (grant 20-02-00505), the Foundation for the Advancement of Theoretical Physics and Mathematics “BASIS”. This work is a contribution of NIST, an agency of the U.S. government, and is not
subject to copyright.
\end{acknowledgments}

\end{document}